\documentclass[fleqn,usenatbib]{mnras}

\usepackage[T1]{fontenc}
\usepackage{newtxtext,newtxmath}
\usepackage{graphicx}
\usepackage{amsmath}
\usepackage{booktabs}
\usepackage{microtype}
\usepackage{fontawesome5}
\newcommand{\rhoD}{\rho_{D}}
\newcommand{\rhoX}{\rho_{X}}
\newcommand{\rhoE}{\rho_{E}}

\title[Multi-centred radial profiles]
{\textsf{RadialPaths}: Radial profiles in multi-centred astronomical structures}

\author[S. de Souza]{Rafael S. de Souza,$^{1,2,3}$\thanks{E-mail: rd23aag@herts.ac.uk}
\thanks{drsouza@ad.unc.edu}\\
$^{1}$Centre for Astrophysics Research, University of Hertfordshire, Hatfield AL10 9AB, UK\\
$^{2}$Instituto de Física, Universidade Federal do Rio Grande do Sul, Porto Alegre, RS 90040-060, Brazil\\
$^{3}$Department of Physics \& Astronomy, University of North Carolina at Chapel Hill, NC 27599-3255, USA
}
\date{Accepted XXX. Received YYY; in original form ZZZ}
\pubyear{2026}

\begin{document}
\label{firstpage}
\pagerange{\pageref{firstpage}--\pageref{lastpage}}
\maketitle

\begin{abstract}
Radial profiles are routinely used to locate structural changes within resolved astronomical systems, but they generally define radial position relative to a single centre. In interacting, clumpy, or otherwise multi-centred structures, the radial location assigned to a feature can depend on the adopted centre. We introduce a centre-conditioned radial representation for connected structures with multiple centres. Distances are measured through the observed source footprint rather than across excluded background, and can be viewed as arrival times of fronts propagating from the centres. Each location is associated with the centre reached first and described by two complementary coordinates: relative centre--boundary depth and normalized progression away from that centre. For an ideal centred circular source, both reduce exactly to the familiar normalized radius $r/R$; in folded, elongated, or tailed geometries they separate because proximity to a lateral boundary and progression through the structure represent different notions of ``outward''. Synthetic examples show that single-centre and equivalent-area radial descriptions do not retain the association between localized features and individual centres, whereas centre-conditioned profiles do. In two JADES DR2 systems, the resulting F277W profiles remain stable under modest changes in the adopted source footprint and one-pixel centre shifts. The construction assumes neither a light-profile law nor an overlapping flux decomposition, and the same radial geometry can be applied to registered maps of continuum light, colour, emission lines, stellar-population properties, and line-of-sight kinematics. An accompanying Python package, \textsf{RadialPaths}, provides the reference implementation together with documentation and online tutorials at \href{https://rafaelsdesouza.com.br/radialpaths}
{GitHub}.

\end{abstract}

\begin{keywords}
galaxies: structure -- galaxies: interactions -- methods: data analysis -- techniques: image processing
\end{keywords}

\section{Introduction}
\label{sec:introduction}

Radial profiles provide one of the most direct descriptions of resolved
astronomical structure. In galaxies, surface-brightness profiles quantify
central concentration and decline, disc structure, breaks and truncations,
rings, and diffuse outer envelopes
\citep{Sersic1963,Freeman1970,PohlenTrujillo2006}. Their use extends well
beyond regular discs: ellipse-based profiles have long been measured for dwarf
and irregular galaxies \citep{SwatersBalcells2002}, while
ultraviolet-to-infrared studies show that the prominence and radial location
of structural components can vary strongly with wavelength
\citep{MunozMateos2009}. Radial colour and stellar-population variations can
therefore connect features in the light distribution to changes in the
underlying stellar populations \citep{Bakos2008}. On sub-galactic scales,
the radial distributions of stellar clumps and their colours, ages, and
star-formation properties are likewise used to investigate clump evolution,
migration, and the growth of their host systems
\citep{Guo2018,Zhu2026}. In each case, interpretation depends not only on the
measured quantity but also on the radial coordinate used to locate it.
Conventionally, that coordinate is defined relative to a single chosen
centre.

That assumption is not always appropriate. Interacting galaxies may contain
two or more bright nuclei within a common envelope, while clumpy galaxies and
stellar complexes may contain several relevant concentrations of light.
Tidal features and irregular outer structure can further make the observed
support strongly non-convex. The resulting difficulty is already visible in
conventional profile analysis. Surface-brightness profiles of merging systems
can become difficult to interpret in the presence of multiple nuclei, and
some analyses restrict profile measurements to nuclei whose morphology
permits a meaningful isophotal description \citep{Rossa2007,Liu2015}.
High-resolution isophotal studies likewise find centre shifts and
position-angle variations associated with neighbouring structures or multiple
nuclei \citep{Goullaud2018}.

Radial measurements nevertheless remain central to current spatially resolved
studies of galaxy interactions. Recent work has used radial profiles of star
formation, molecular-gas properties, and metallicity to trace how
merger-driven changes vary across galaxies and with interaction stage
\citep{Thorp2024,GaraySolis2025,GaraySolis2026}. A particularly direct
illustration of the limitation considered here comes from resolved
post-merger maps, in which star formation can be clumpy, offset from the
centre, and associated with non-axisymmetric tidal features, so that a single
radial profile does not retain the full spatial information
\citep{Thorp2019}. This ambiguity is also apparent in spatially resolved JWST studies,
where complex morphologies can make individual radial gradients difficult
to interpret \citep{Tripodi2024}.
In such systems, adopting one global centre can mix
structure associated with different concentrations of light.
Moreover, a circular or elliptical radius is defined in the image plane and can connect locations across regions outside the observed structure, so locations
assigned the same nominal radius need not occupy comparable positions within
the connected source. The problem therefore arises before deciding whether a
S\'ersic, exponential, broken, or other profile law provides a good
description: the meaning of radial position itself has become ambiguous.

Existing radial-profile methods relax different parts of this problem.
Elliptical-isophote methods \citep{Jedrzejewski1987} and pipelines such as
\textsc{Isofit} \citep{Ciambur2015} and \textsc{AutoProf}
\citep{Stone2021AutoProf} allow the fitted isophotal geometry and higher-order
structure to vary with radius, but still produce a single isophotal radial
sequence rather than simultaneous centre-conditioned orderings for multiple
centres.
Equivalent-radius profiles instead characterize strongly irregular systems
through the area enclosed at successive surface-brightness levels
\citep[e.g.][]{Papaderos1996,Cairos2001,Micheva2013}. This accommodates
arbitrarily shaped, even disconnected isophotal regions without imposing an
elliptical geometry. The resulting profile nevertheless assigns each level a
single equivalent radius and does not retain the identity of multiple
centres. Because surface brightness itself defines the ordering, it also addresses a different question from a radial coordinate whose ordering is geometric once the support and centres are specified.

Other methods operate once a radial ordering or structural representation has
already been chosen. One-dimensional tools such as \textsc{Profiler} model an
observed radial profile with flexible combinations of analytic components
\citep{Ciambur2016}, while derivatives or curvature can expose transitions
within an already ordered profile \citep{LucatelliFerrari2019}. Parametric
two-dimensional decompositions, including \textsc{GALFIT}
\citep{Peng2002,Peng2010}, \textsc{IMFIT} \citep{Erwin2015}, and
\textsc{Galmoss} \citep{Chen2024Galmoss}, instead ask which combination of
overlapping light components can reproduce the image and provide quantities
such as component fluxes, sizes, shapes, and S\'ersic indices.
Non-parametric morphology frameworks such as \textsc{Morfometryka}
\citep{Ferrari2015} and \textsc{statmorph}
\citep{RodriguezGomez2019} summarize other aspects of structural complexity,
including concentration, asymmetry, smoothness, Gini--$M_{20}$, and
multimodality \citep{Conselice2003,Lotz2004,Freeman2013}. These approaches
address complementary scientific questions, but do not by themselves define
a radial ordering that both retains the identity of multiple centres and
remains constrained to the connected structure. The question considered here
is therefore one of radial localization: given several relevant centres
within one connected structure, where does an observed feature occur relative
to its associated centre and to the geometry of the structure itself?

Here we address this problem by defining a centre-conditioned radial
representation for a connected astronomical structure containing multiple centres. We use ``centre'' operationally throughout: a centre is a reference point specified for radial profiling and need not be a dynamical centre, a confirmed nucleus, or an independently identified physical component.
Distances constrained to remain within the observed support first associate
each location with one of these centres. Within each centre-associated region, we then distinguish two complementary
notions of radial position: relative centre--boundary depth and normalized
progression away from the centre. These coordinates answer different geometric questions rather
than provide competing estimates of a single radius. Both reduce to
the conventional normalized radius for a centred circular source, while
differing naturally in folded, elongated, or tailed structures. The resulting
profiles retain centre identity and the radial location of breaks, shoulders,
extrema, truncations, and other localized structure without requiring a
particular light-profile law.

We examine what information is retained relative to single-centre and
equivalent-area radial descriptions, isolate the behaviour of the two coordinates in
controlled non-convex geometries, and illustrate the construction with two
multi-centred systems in JADES DR2 imaging. We also test the sensitivity of
the resulting centre assignments and profiles to modest changes in the
adopted support and  centre positions.

The remainder of this paper is organized as follows.
Section~\ref{sec:methodology} introduces the multi-centre radial construction,
its geometric motivation, and its relation to established profile and
decomposition methods. Section~\ref{sec:controlled} examines the behaviour
of the proposed coordinates in compact and non-convex geometries.
Section~\ref{sec:data} describes the JADES imaging and the two illustrative
systems, and Section~\ref{sec:jades} applies the construction to their
resolved light distributions. Finally, Section~\ref{sec:conclusions}
discusses the interpretation and limitations of the framework and summarizes
the main conclusions.

\section{Methodology}
\label{sec:methodology}

\subsection{Geometric setting and relation to established methods}
\label{sec:context}

When more than one centre is relevant, the main ambiguity is not which
profile law best describes the observed field, but what ``radial position''
should mean in the first place. A useful construction should follow the
observed structure rather than measure distance across excluded background,
retain the identity of the centre relative to which a position is defined,
and reduce to the familiar normalized radius in the centred circular limit.
It should also be dimensionless and, before discretization on a pixel grid,
unchanged by translation, rotation, or uniform rescaling. Irregular
structures introduce one further complication: being close to the local
boundary is not the same as having progressed far through the structure. A
point in a narrow arm, for example, may already be boundary-like while still
lying only partway along the arm. We therefore keep these two notions of
``outward'' separate.

The geometry needed to do this is familiar. Voronoi partitions assign
locations to their nearest centre \citep{Voronoi1908}, while shortest-path
distances on graphs provide a standard way to measure distance when only
certain routes are admissible \citep{Dijkstra1959}. Geodesic Voronoi diagrams
combine these ideas inside a bounded domain \citep{Aronov1987}. Related geodesic constructions appear in seeded watershed methods
\citep{VincentSoille1991}, while object-level radial measurements from a centre to the boundary of an irregular segmented region are implemented in biological image-analysis tools such as \textsc{CellProfiler}
\citep{Carpenter2006,CellProfilerRadial}. Voronoi tessellations have also been
used to construct radial number-density profiles for discrete astronomical
populations \citep{DornanHarris2024}. That approach addresses the spatial
distribution of point-like tracers rather than surface-brightness fields
defined over a connected image footprint.

Existing astronomical profile methods relax different parts of the problem.
Isophotal approaches such as \textsc{Isofit} \citep{Ciambur2015} and
\textsc{AutoProf} \citep{Stone2021AutoProf} allow ellipticity, position angle,
and higher-order departures from elliptical isophotes to vary with radius.
They therefore retain substantially more information about the changing shape
of the source than profiles measured in fixed circular or elliptical annuli.
The result, however, is still a single isophotal sequence through the galaxy,
rather than simultaneous radial orderings tied to several centres.

Equivalent-radius profiles take a different route
\citep[e.g.][]{Papaderos1996,Cairos2001,Micheva2013}. At a given
surface-brightness level, the contributing pixels may form a highly irregular
or even disconnected region; mapping their combined area to a circle of equal
area provides a scalar equivalent radius without fitting an ellipse. This is
particularly useful for irregular galaxies, but each surface-brightness level
is still represented by a single radial coordinate and the identity of several
centres is not retained. \textsc{Profiler} \citep{Ciambur2016}, in turn,
addresses a downstream modelling problem: it fits an already-defined radial
surface-brightness profile with S\'ersic, exponential, and other analytic
components.

Our question comes one step earlier. We want to define radial position itself
when more than one centre is relevant. Our contribution is not a new distance
or partitioning algorithm, but the combination of support-constrained distance
with nearest-centre partitioning to define centre-conditioned radial position
within a connected astronomical structure. Once the connected support and
centres are specified, locations are assigned geometrically rather than
ordered by the value of the tracer being measured. This produces a separate
radial description for each centre without decomposing the image into
overlapping light components.

\begin{figure*}
 \centering
 \includegraphics[width=\textwidth]{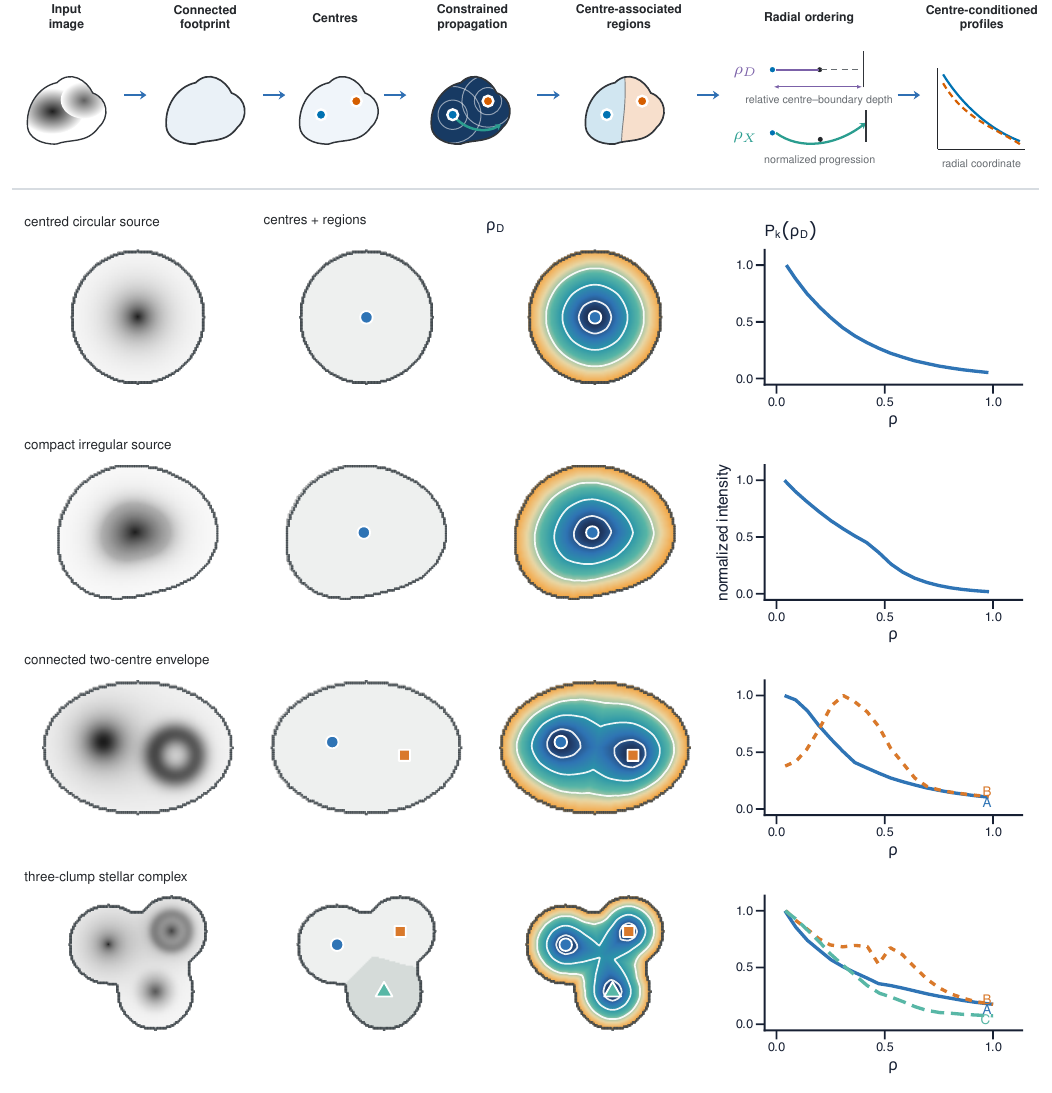}
\caption{\textbf{Top:} conceptual workflow of centre-conditioned radial
profiling. Starting from an observed intensity field, a connected source
footprint and a set of centres define the geometry. Constrained propagation
within the footprint associates each location with the centre reached first,
producing centre-associated regions. Radial position is then described by the
relative centre--boundary depth $\rhoD$ or the normalized progression $\rhoX$,
and the measured field is summarized as a separate profile for each centre.
\textbf{Bottom:} the same construction illustrated for a centred circular
source, a compact irregular source, a connected two-centre envelope, and a
three-clump stellar complex. The lower examples show $\rhoD$ and its
corresponding profiles. The circular case recovers the conventional normalized
radial ordering, whereas the irregular examples follow paths confined to the
source footprint. In the two-centre case, a ring-like feature associated with
Centre B remains localized in its corresponding profile rather than being
combined with structure around Centre A. The final example shows that the
construction extends naturally to more than two centres.}
 \label{fig:construction}
\end{figure*}

Figure~\ref{fig:construction} summarizes the construction visually. The top
strip shows the workflow from an observed field, source footprint, and centres
to constrained propagation, centre-associated regions, radial ordering, and
centre-conditioned profiles. The lower examples illustrate the same
construction in increasingly complex geometries.

In the first lower row, a single centre in a circular source footprint
recovers the familiar normalized radial ordering. The second lower row retains
one centre but removes circular symmetry: distances follow admissible paths
inside the footprint, so the radial ordering does not cut across excluded
background. The third lower row introduces the multi-centre case. Two centres
occupy a common connected envelope, and each location is associated with the
centre reached first. A ring-like feature around Centre B therefore remains
localized in the profile of Centre B rather than being combined with structure
around Centre A. The final lower row extends the same construction to three
centres, showing how shoulders, truncations, and other localized features
remain tied to the relevant centre in a more complex system.

The distinction between geometry and tracer is also visible in the figure.
Once the support and centres are fixed, the intensity field determines what
is measured in the profiles, but not the radial ordering itself. The lower
examples visualize only the relative centre--boundary depth $\rhoD$, while
the workflow introduces both coordinates; their different behaviour in
irregular structures is examined in Section~\ref{sec:controlled}.

\subsection{Centre-associated regions and radial coordinates}
\label{sec:coordinates}

We now formalize the construction previewed in
Fig.~\ref{fig:construction}. Let $\Omega\subset\mathbb{R}^{2}$ denote a
bounded, path-connected source
footprint, which we refer to below as the support, and let
$\mathcal{C}=\{c_1,\ldots,c_K\}\subset\operatorname{int}(\Omega)$ denote the set of reference centres. Distances are measured along paths that remain inside
$\Omega$. Two locations that are close in the surrounding image plane may
therefore be far apart within the observed structure if a direct path between
them crosses excluded background.
Let $d_G$ denote this support-constrained shortest-path distance. The distance
from centre $c_k$ to a location $x$ is

\begin{equation}
 d_k(x)=d_G(c_k,x).
 \label{eq:centre_distance}
\end{equation}

A useful way to picture $d_k$ is to imagine a front launched from $c_k$ and propagating at unit speed through the accepted support. Its arrival time at $x$ is $d_k(x)$. If fronts are launched simultaneously from all centres, each location is
associated with the centre whose front arrives first. We resolve an exact tie by choosing the lowest-index centre,
\begin{equation}
 a(x)=
 \min\left\{
 j:\,
 d_j(x)=\min_{\ell}d_\ell(x)
 \right\},
 \label{eq:assignment}
\end{equation}
and define the centre-associated region
\begin{equation}
 B_k=\{x\in\Omega:a(x)=k\}.
 \label{eq:centre_region}
\end{equation}
The regions $B_k$ partition the connected support according to nearest-centre
distance measured through the structure. Once the support and centres have been fixed, this assignment is geometric: neither the value nor
the local gradient of the tracer being profiled enters the partition. The construction is therefore a geodesic Voronoi partition rather than a
watershed segmentation, although the resulting regions may look superficially similar. In the front-propagation picture, the interfaces between neighbouring
regions are simply the locations where fronts from different centres arrive at the same time.

Once a location has been assigned to a centre, two different radial questions remain. One asks where the location lies between that centre and the nearest edge of the accepted structure. The other asks how far it has progressed away
from the centre relative to the full extent of its centre-associated region. These questions give the same answer in the familiar centred circular case, but not in an irregular support.

For $x\in B_k$, define the distance to the nearest boundary of the accepted support as
\begin{equation}
 b(x)=d_G(x,\partial\Omega)
     \equiv \inf_{y\in\partial\Omega}d_G(x,y).
 \label{eq:boundary_distance}
\end{equation}
Here $\partial\Omega$ includes both the exterior boundary and the boundary of
any internal excluded region. It does not include the interface between neighbouring centre-associated regions. We define the radial extent of $B_k$ as

\begin{equation}
 L_k=\max_{x\in B_k}d_k(x).
 \label{eq:radial_extent}
\end{equation}
For the non-degenerate regions considered here, $L_k>0$.
The first radial coordinate compares the distance from the centre with the distance to the nearest support boundary,
\begin{equation}
 \rho_{D,k}(x)
 =
 \frac{d_k(x)}
      {d_k(x)+b(x)}.
 \label{eq:rhoD}
\end{equation}

We call this the \emph{relative centre--boundary depth}. It is zero at the  centre and increases as a location becomes boundary-like relative to
its distance from that centre. In a narrow arm or tail, $\rho_{D,k}$ can therefore become large because a lateral boundary is nearby, even when the location has travelled only partway along the structure.

The second coordinate instead normalizes the centre distance by the radial extent of the same centre-associated region,
\begin{equation}
 \rho_{X,k}(x)
 =
 \frac{d_k(x)}{L_k}.
 \label{eq:rhoX}
\end{equation}
We call this \emph{normalized progression}. It is zero at the centre and reaches unity at locations attaining the radial extent of the region. Unlike
$\rho_{D,k}$, it does not respond directly to the distance from a lateral boundary. Along an elongated arm or tail, it can therefore provide a longitudinal ordering: a narrow section may already have a large $\rhoD$ while remaining at an intermediate value of $\rhoX$.

The front-propagation picture makes the distinction especially direct. The distance $d_k(x)$ is the arrival time of a front travelling outward from the 
centre, while $b(x)$ is the arrival time of a front travelling inward from the nearest support boundary. The coordinate $\rhoD$ compares these two arrival times and therefore describes depth within the local geometry.
The coordinate $\rhoX$ instead compares the centre-front arrival time with the latest arrival within the same centre-associated region, and therefore describes progression away from the centre. Appendix~\ref{app:distance}
develops this interpretation and describes the numerical evaluation of $d_G$.

The two coordinates are thus complementary descriptions of radial position, not competing estimates of a single underlying radius. Both are dimensionless and, before discretization on a pixel grid, unchanged by translation, rotation, or uniform rescaling. They are not intended to provide
a unique coordinate system for an irregular structure. Rather, they retain two useful meanings of ``outward'' that coincide in the conventional symmetric limit.

For a circular support of radius $R$ with one centre at its geometric centre,
\begin{equation}
 d_1=r,\qquad
 b=R-r,\qquad
 L_1=R,
 \label{eq:circular_distances}
\end{equation}
and hence
\begin{equation}
 \rho_D=\rho_X=\frac{r}{R}.
 \label{eq:circular_limit}
\end{equation}
Both coordinates therefore recover the conventional normalized radius
exactly for an ideal centred circular source. On a finite pixel grid, the support-constrained distances inherit the usual lattice approximation, so rotational invariance is preserved only approximately. The size of this discretization effect is quantified in Appendix~\ref{app:distance}.

\subsection{Centre-conditioned profiles}
\label{sec:profiles}

Once the support and  centres have fixed the radial geometry,
constructing a profile is a separate and familiar step. The geometry
determines which centre a location belongs to and where it lies under
$\rhoD$ or $\rhoX$; it makes no assumption about how the measured field
varies with radial position. For the profiles used in this paper, a bin
centred on $\rho$ with width $\Delta\rho$ is summarized by the median,

\begin{equation}
\begin{split}
P_k(\rho)
&=
\operatorname{median}
\Bigl\{
I(x):
x\in B_k,
\\[-0.25ex]
&\qquad
\rho-\Delta\rho/{2}
\leq \rho_k(x)
<
\rho+\Delta\rho/{2}
\Bigr\}.
\end{split}
\label{eq:profile}
\end{equation}

where $\rho_k$ denotes either $\rho_{D,k}$ or $\rho_{X,k}$. For a finite
binning of $[0,1]$, the upper endpoint is included in the final bin so that
locations with $\rho_k=1$ are retained.

The values within a bin need not be tightly concentrated. Their variation can
be summarized by the 16th--84th percentile interval, as in the examples
below, or retained as a full within-bin distribution when the spatial scatter
is itself of interest. A smooth function $F_k(\rho)$ may subsequently be
fitted to the binned measurements, for example with a penalized spline, but
such smoothing does not enter the centre assignment, radial coordinates, or
profile definition.

For either radial coordinate, the result is a family of profiles,
one for each centre, rather than a single curve obtained
by pooling the full structure. Retaining the centre
label is essential: combining locations from different centre-associated
regions at the same value of $\rho$ would mix structures anchored to
different reference points. A localized break, shoulder, extremum,
truncation, or other transition can instead be tied to its associated centre
and described by its positions under both $\rhoD$ and $\rhoX$, together with
an appropriate amplitude or shape summary.

The geometry can also be reused when the measured tracer changes. For any
registered scalar field $Q(x)$ defined on the same support, replacing $I(x)$
by $Q(x)$ in Equation~(\ref{eq:profile}) changes the values summarized in
each radial bin but not the centre assignment or radial ordering. Possible
examples include colour, emission-line intensity or equivalent width,
resolved stellar-population or dust properties, and line-of-sight velocity
or velocity dispersion. The field need not be non-negative; signed quantities
such as velocity can be profiled in the same way. Meaningful comparisons between tracers require compatible registration and
masking and, where relevant, matched or explicitly modelled spatial
resolution. The astronomical examples below use F277W intensity, but no change to the radial construction is required for another registered scalar field.

\subsection{Relation to additive image decomposition}
\label{sec:decomposition}

Centre-conditioned radial profiles and multi-component image decomposition
answer different questions. A parametric decomposition represents the
observed field as a superposition of components, schematically

\begin{equation}
 I(x)\simeq \sum_m S_m(x\mid\theta_m),
 \label{eq:decomposition}
\end{equation}

where several components may contribute light at the same image location and
the parameters $\theta_m$ describe quantities such as flux, scale radius,
shape, or S\'ersic index. The aim is to determine which combination of
overlapping components reproduces the observed image.

The construction defined above does not assign the observed light among
components. Instead, it asks where the measured field lies radially relative
to each centre once the connected support and centres have been fixed. A
location belongs to one centre-associated region for the purpose of radial
ordering, but this does not imply that its flux originates from a single
physical component. The partition is geometric, not a decomposition of the
light.

This distinction is especially clear in the conventional single-centre case.
For a centred circular source, $\rhoD=\rhoX=r/R$, so the usual S\'ersic,
exponential, broken, or non-parametric profile descriptions can be applied
directly after the radial coordinate has been defined. The same logic carries
over to a centre-conditioned profile: an analytic model may be fitted when
useful for a particular scientific question, but no profile law is required
to construct the radial ordering itself. Radial profiling and additive decomposition are therefore complementary
approaches rather than competing descriptions of the same quantity.

\subsection{Information retained by different radial descriptions}
\label{sec:retained_information}

\begin{figure*}
 \centering
 \includegraphics[width=\textwidth]{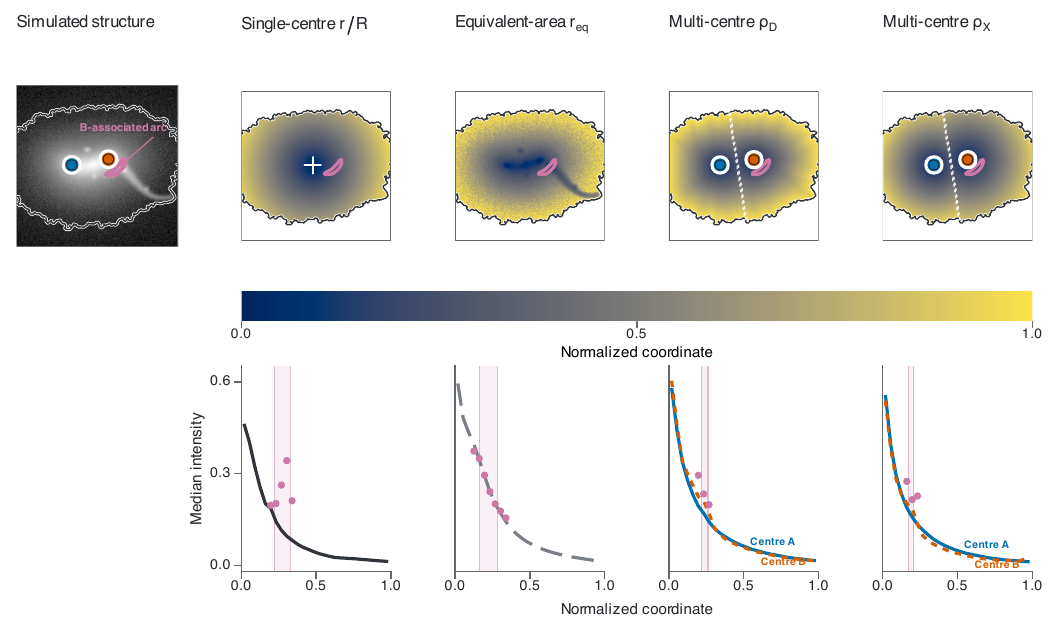}
\caption{Information retained by different radial descriptions of the same
simulated astronomical structure. The source contains two unequal light
concentrations within a common irregular envelope, compact substructure, and
a low-contrast arc associated by construction with Centre B. To the right of
the source image, four columns compare a conventional single-centre normalized
radius $r/R$, an intensity-ranked equivalent-area radius $r_{\rm eq}$, and the
centre-conditioned coordinates $\rhoD$ and $\rhoX$. In each comparison
column, the upper panel shows the corresponding radial coordinate field and
the lower panel the profile measured from the same intensity field. Magenta
points and intervals mark the radial locations occupied by pixels belonging
to the injected arc. The single-centre and equivalent-area descriptions
assign the feature a radial location but do not retain its association with
Centre B, whereas the centre-conditioned profiles retain both centre identity and radial location.}
\label{fig:information}
\end{figure*}

Before examining $\rhoD$ and $\rhoX$ separately, we ask what happens when the
same multi-centred intensity field is described using different
one-dimensional radial coordinates. Figure~\ref{fig:information} compares the
centre-conditioned coordinates with two global radial descriptions: a
conventional single-centre normalized radius and an equivalent-area radius.

For the single-centre baseline, we use the positive background-subtracted
flux centroid of the accepted support as the reference point. The Euclidean
distance $r$ from this point is normalized by its maximum value $R$ within
the support, giving $r/R$. This provides a valid global radial description,
but every location is necessarily referred to the same centre.

For the equivalent-area baseline, we order accepted-support pixels by surface
brightness. If $A$ is the cumulative area associated with a given intensity
rank and $A_\Omega$ is the total area of the support, we define

\begin{equation}
 \frac{r_{\rm eq}}{r_{{\rm eq},\max}}
 =
 \sqrt{\frac{A}{A_\Omega}},
 \label{eq:req}
\end{equation}
equivalent to $r_{\rm eq}=\sqrt{A/\pi}$ up to normalization by the total support area. This description accommodates an arbitrarily irregular
collection of pixels without fitting circular or elliptical isophotes, but the ordering is determined by surface brightness and is ultimately expressed through a single scalar coordinate.

The simulated structure in Fig.~\ref{fig:information} was constructed independently of $\rhoD$ and $\rhoX$. It contains two unequal light
concentrations within one connected envelope, localized substructure, and a low-contrast arc associated by construction with Centre B. These features allow us to ask not which radial description is
``best'', but what information each description retains.
The distinction is clearest for the injected arc. Both global descriptions
assign it a radial location, but neither retains the fact that the feature is
associated with Centre B. The centre-conditioned profiles retain both pieces of information: where the feature lies radially and which centre provides the corresponding reference.

For this synthetic comparison, support-constrained distances are evaluated
with a fast-marching approximation that propagates distance from each centre
without leaving the source footprint. The relation between this numerical
implementation and the graph-based distances used elsewhere is discussed in
Appendix~\ref{app:distance}.

\section{Radial coordinates in irregular structures}
\label{sec:controlled}

\begin{figure*}
 \centering
 \includegraphics[width=\textwidth]{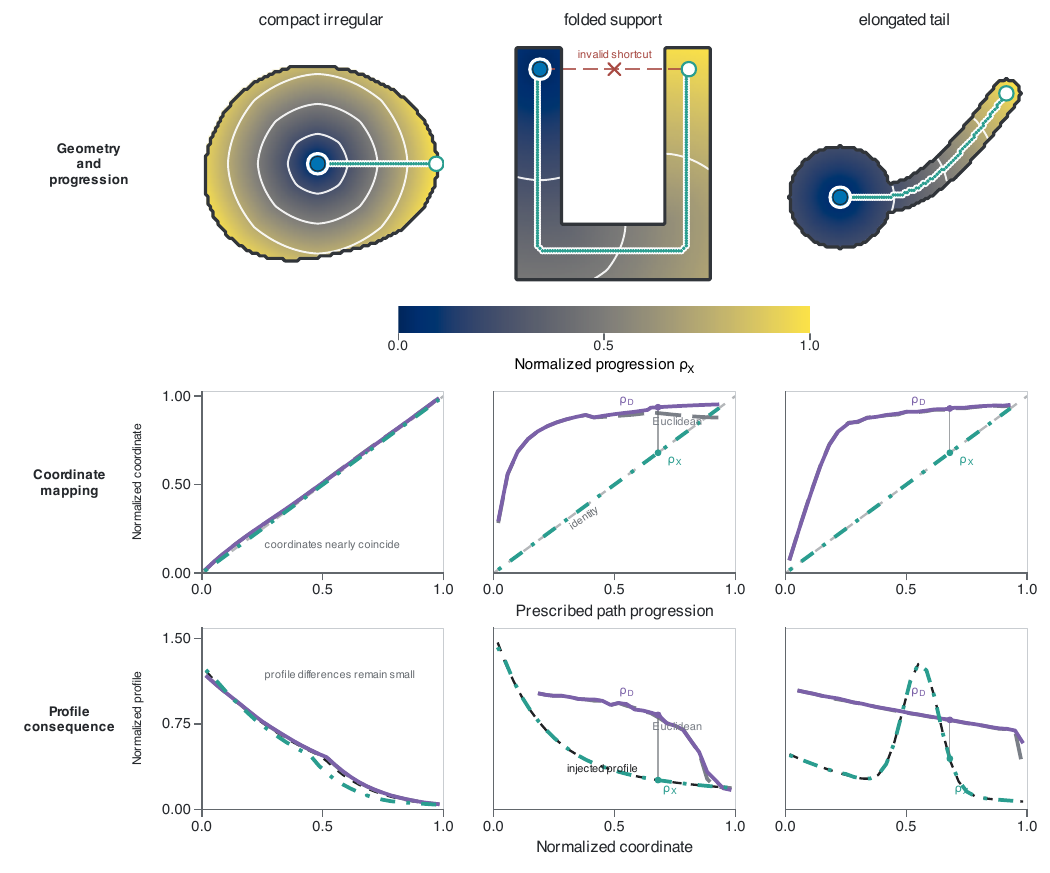}
 \caption{Behaviour of the radial coordinates in compact, folded, and
 elongated supports. Columns show the three geometries, while rows show the
 geometry and prescribed progression, the corresponding coordinate mappings,
 and the resulting profiles. Top: normalized progression $\rhoX$, selected
 contours, and the prescribed in-support path. In the folded support, the
 dashed red segment marks a straight-line shortcut that crosses excluded
 background and is therefore not admissible under the support-constrained
 distance. Middle: Euclidean depth $\rhoE$ (grey), relative centre--boundary
 depth $\rhoD$ (purple), and normalized progression $\rhoX$ (teal) as
 functions of the prescribed path progression; the pale dashed line marks
 identity. Bottom: profiles recovered from the same scalar field, with the
 injected profile shown for reference. The coordinates give nearly the same
 ordering in the compact case, whereas folding separates Euclidean from
 in-support distance and lateral narrowing separates boundary depth from
 progression.}
 \label{fig:geometry}
\end{figure*}

We next isolate the effect of geometry from observational complications. We
consider three controlled supports: a compact irregular source, a folded
U-shaped structure, and an elongated curved tail. In each case, progression
through the connected structure is prescribed independently of $\rhoD$ and
$\rhoX$. The associated scalar fields contain smooth radial variation
together with localized breaks, rings, truncations, or outer enhancements;
none of these features is defined in terms of the proposed coordinates.

As a reference, we also define a Euclidean depth coordinate,

\begin{equation}
 \rho_E=\frac{d_E}{d_E+b_E},
 \label{eq:rhoE}
\end{equation}

where $d_E$ and $b_E$ are the straight-line distances to the centre and to the
support boundary, respectively. Unlike the support-constrained coordinates,
$\rho_E$ does not require the connecting path to remain inside the observed
structure. It can therefore assign similar radial positions to locations that
are separated by excluded background.

Figure~\ref{fig:geometry} shows when these differences matter. In the compact
irregular support, Euclidean depth, relative centre--boundary depth, and
normalized progression remain close to one another. The geometry is
sufficiently compact that straight-line and in-support distances give almost
the same radial ordering, and the corresponding profiles remain similar.

The folded support isolates the effect of an inadmissible shortcut. Locations
on opposite sides of the fold can lie close together in the image plane even
though any path confined to the support must travel around the U shape. The
Euclidean coordinate therefore shortens the apparent separation, whereas the
support-constrained coordinates follow the connected structure. Along the
prescribed path, $\rhoX$ continues to track progression through the fold,
while $\rhoD$ also responds as locations approach a nearby lateral boundary.

The elongated tail separates $\rhoD$ and $\rhoX$ even without a fold. As the
tail narrows, a location can become close to the boundary while still lying
only partway along the feature. The same location can therefore have a large
$\rhoD$ but an intermediate $\rhoX$. The resulting profiles differ because
the intensity field is being ordered according to two genuinely different
notions of radial position.

These examples show that the coordinates should not be interpreted as
alternative estimates of a single underlying radius. For compact geometries
they can produce almost the same ordering, but folds and lateral narrowing
separate them in predictable ways. Which radial description is most useful
therefore depends on the feature being localized: whether the relevant notion
of ``outward'' is proximity to the local boundary or progression away from
the centre through the connected structure.

\section{Illustrative galaxy sample}
\label{sec:data}

We use public JWST/NIRCam imaging from the JADES survey in GOODS-S
\citep{Rieke2023,Eisenstein2025}. The data resolve compact light
concentrations, clumps, tidal structure, and irregular outer emission on the
same image, making them well suited to cases in which a single global centre
is not an adequate radial reference.

The two systems shown here were identified from the Galaxy Zoo: CANDELS
catalogue \citep{Simmons2017} as illustrative cases containing two prominent
light concentrations within a common irregular structure. They are not
intended to define a representative merger sample, and we do not infer a
dynamical state or physical association from morphology alone. The centres
used below therefore serve only as reference points for radial profiling.

For each system, we define the reference geometry from the native JADES
F277W image, with a pixel scale of approximately $0.030$ arcsec
pixel$^{-1}$. We estimate the local background iteratively from pixels outside
a provisional source region and define
\begin{equation}
 \sigma_{\rm bg}=1.4826\,{\rm MAD},
\end{equation}
where the MAD is measured about the background median. The image is then
smoothed with a Gaussian matched to the empirical F277W PSF,
$0.1123$ arcsec FWHM, or $3.743$ native pixels. The initial support contains
pixels satisfying
\begin{equation}
 I_{\rm smooth}\geq I_{\rm bg}+1.5\sigma_{\rm bg}.
 \label{eq:support_threshold}
\end{equation}
We apply binary closing followed by dilation with a circular structuring
element of radius $0.5$ PSF FWHM. Internal excluded regions are left unfilled,
so their edges remain part of the support boundary.
Centres are identified from local maxima within a $9\times9$ pixel neighbourhood of the same smoothed F277W image, corresponding to a half-width of approximately one PSF FWHM.
Candidate maxima must satisfy
\begin{equation}
 I_{\rm smooth}\geq I_{\rm bg}+5\sigma_{\rm bg}.
 \label{eq:centre_threshold}
\end{equation}
The two brightest accepted maxima are adopted as the centres, provided that
they are separated by at least two PSF FWHM. We then retain the
eight-connected component containing both centres and use this region as the
support for the radial construction.

Distances are evaluated on the native image grid. We represent the support
as an eight-neighbour graph, with axial edge length unity and diagonal edge
length $\sqrt{2}$, and evaluate shortest paths without allowing them to cross
excluded pixels.
Because the JADES mosaics are astrometrically registered, the same support and
centres can also be used in other NIRCam bands when a common radial geometry
is desired. Any quantitative comparison between bands must still account for
differences in PSF and spatial resolution.

\section{Application to two JADES galaxies}
\label{sec:jades}

\begin{figure*}
 \centering
 \includegraphics[width=\textwidth]{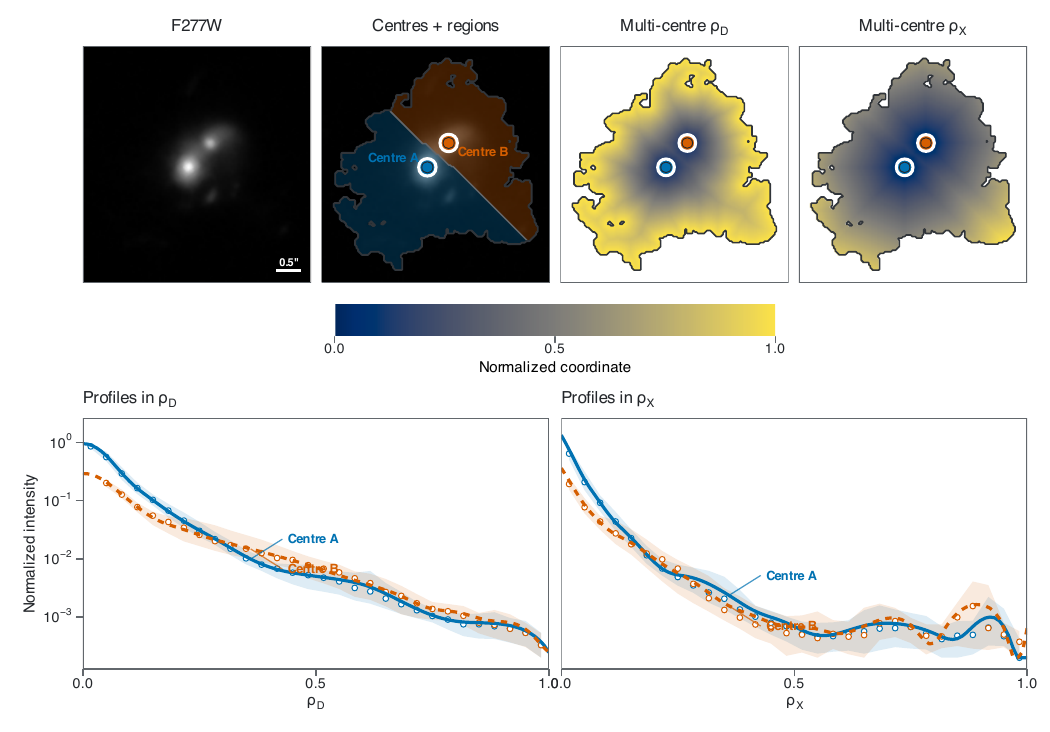}
 \caption{First JADES DR2 example, GZMERGER09 (Galaxy Zoo: CANDELS
 GDS\_2859; $\alpha_{\rm J2000}=53.102300^\circ$,
 $\delta_{\rm J2000}=-27.887726^\circ$). Top: observed NIRCam F277W image,
 the two centre-associated regions, and the corresponding $\rhoD$ and $\rhoX$
 coordinate fields. Bottom: binned median profiles of the
 background-subtracted F277W intensity, normalized by the maximum intensity
 within the adopted support, with 16--84 percentile ranges for the two
 centres. Smooth curves are included only as visual summaries. Equal
 numerical values of $\rhoD$ and $\rhoX$ have different geometric meanings:
 relative centre--boundary depth and normalized progression, respectively.
 The coordinate fields are interpolated only for display; all profile
 measurements are evaluated on the native image grid.}
 \label{fig:realcase}
\end{figure*}

\begin{figure*}
 \centering
 \includegraphics[width=0.90\textwidth]{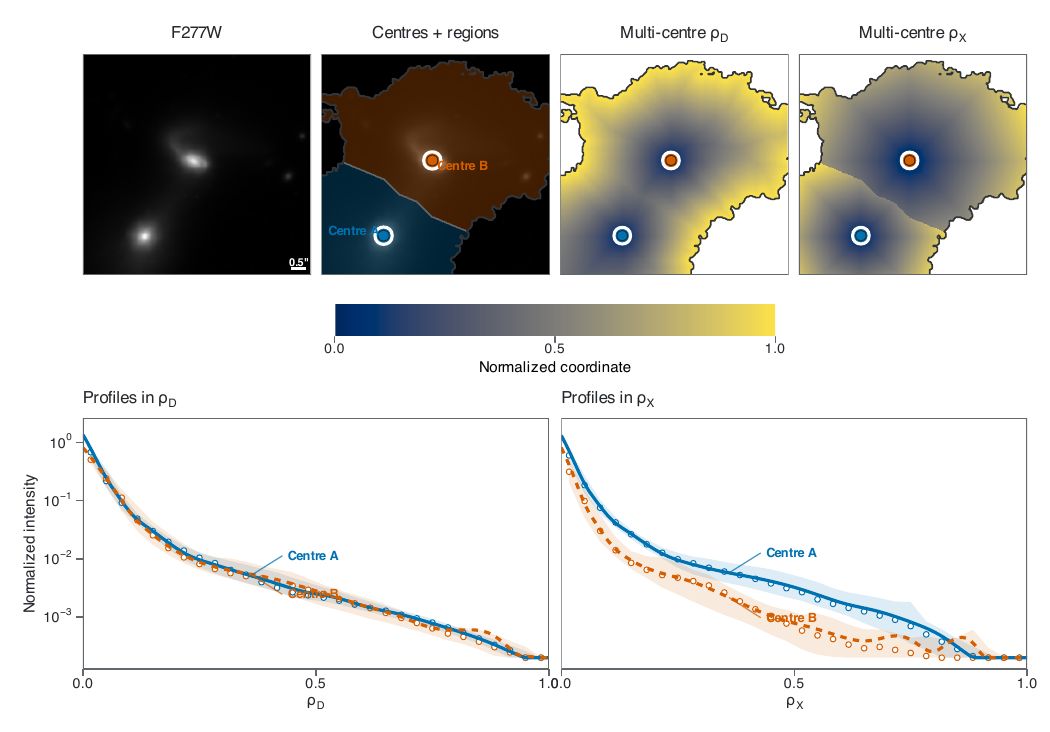}
 \caption{Second JADES DR2 example, GZMERGER29 (Galaxy Zoo: CANDELS
 GDS\_4608; $\alpha_{\rm J2000}=53.081307^\circ$,
 $\delta_{\rm J2000}=-27.871586^\circ$), shown using the same representation
 as Fig.~\ref{fig:realcase}. The two centre-associated regions have different
 outer geometries. Their normalized F277W intensity profiles can therefore be
 similar at a given relative centre--boundary depth while differing in their
 progression away from the corresponding centres.}
 \label{fig:realcase29}
\end{figure*}

For each centre, we construct F277W intensity profiles using both radial
coordinates defined in Section~\ref{sec:coordinates}. We use 30 equal bins
over $[0,1]$, corresponding to $\Delta\rho=1/30$, and omit bins containing
fewer than six native pixels. Each bin is summarized by the unweighted median
of the background-subtracted pixel intensities. The 16--84 percentile
intervals show the variation of the intensity within the bin, rather than the
uncertainty on the median. For Figs.~\ref{fig:realcase} and
\ref{fig:realcase29}, the intensities are normalized by the maximum
background-subtracted intensity within the adopted support.

The smooth curves shown with the binned profiles are included only as visual
summaries. We fit a penalized cubic B-spline with 18 coefficients and a fixed
second-difference penalty of 10. The spline
does not enter the centre assignment or the radial coordinates and is not
interpreted as an assumed light-profile law.

Figure~\ref{fig:realcase} shows the first galaxy. Its two centres lie in
different parts of the same irregular envelope. A single global radius would
place locations associated with both light concentrations on one radial axis.
The centre-conditioned construction instead keeps their identities separate,
so the radial variation of the observed light can be followed independently
within the two centre-associated regions.

The two coordinates then emphasize different aspects of that same light
distribution. The depth coordinate $\rhoD$ orders the surface brightness
according to the balance between distance from the centre and proximity to
the local boundary. The progression coordinate $\rhoX$ instead measures how
far a location lies from the centre relative to the full radial extent of its
region. A feature can therefore occur at a similar relative depth in the two
regions while lying at different stages of progression through them. This
difference comes from the geometry itself; it does not require the light to be
decomposed into overlapping S\'ersic or other parametric components.

Figure~\ref{fig:realcase29} shows a second geometry. Here the two
centre-associated regions differ more strongly in their outer shape. Their
profiles in $\rhoD$ remain similar over much of their relative depth, while
the $\rhoX$ profiles separate more clearly as one moves away from the
centres. Similarity at a given centre--boundary depth therefore does not imply
similarity at the same stage of progression through the connected structure.
This is the same distinction isolated in the synthetic examples of
Section~\ref{sec:controlled}, now appearing in an observed galaxy.

We also tested how sensitive the profiles are to the adopted support and to
small changes in the centre positions. Repeating the construction with
support thresholds of $1.25\sigma_{\rm bg}$ and $1.75\sigma_{\rm bg}$ changed
no centre assignments among pixels shared by the reference and perturbed
supports. Shifting either centre by one native pixel in each of the eight
neighbouring directions changed at most 1.53 per cent of the assignments for
GZMERGER09 and 0.23 per cent for GZMERGER29.

The radial profiles themselves were similarly stable. Every bin populated in
the reference profile remained populated under all perturbations. Relative to
the 5--95 percentile dynamic range of the reference profile, the largest RMS
difference was 0.060 for the support-threshold perturbations and 0.137 for the
one-pixel centre shifts. Excluding the innermost radial bin reduced these
values to 0.016 and 0.082, respectively. Across the centre shifts, the largest
median absolute bin-by-bin difference was only 0.002 of the reference dynamic
range, indicating that the larger deviations are confined to a small number
of radial bins rather than extending across the full profile.

All profiles shown here are measured directly from the observed images and
therefore retain the effects of the instrumental point-spread function.
Physical comparisons between bands, or between observations with different
spatial resolution, would require PSF matching or explicit forward modelling.
The construction determines where a feature lies relative to a centre; it
does not determine what that centre represents physically.

\section{Discussion and conclusions}
\label{sec:conclusions}

Radial profiles are useful because they attach structural changes to a
position within an astronomical system. Breaks and truncations in discs,
changes in colour or stellar populations, rings, and extended envelopes all
derive part of their interpretation from where they occur relative to the
system under study
\citep{PohlenTrujillo2006,Bakos2008,MunozMateos2009}. The main result of
this work is that this radial localization can be retained even when no single
centre provides an adequate ordering of the structure. Rather than collapsing
a multi-centred system onto one global radial coordinate, the construction
retains centre identity and describes the observed field separately within
the corresponding centre-associated regions.

The two coordinates preserve complementary notions of radial position. The
depth coordinate $\rhoD$ measures the balance between distance from a centre
and proximity to the local boundary, while the progression coordinate $\rhoX$
measures distance from the centre relative to the radial extent of its
centre-associated region. They coincide with the conventional normalized
radius for a centred circular source but separate naturally in irregular,
folded, or elongated structures. A location in a narrow arm or tidal feature,
for example, may already lie close to a lateral boundary while remaining only
partway through the structure. The controlled examples show that this
difference reflects two distinct geometric meanings of radial position rather
than disagreement between alternative estimates of the same radius. Which
coordinate is more useful therefore depends on the feature being localized.

This also clarifies the distinction from parametric decomposition. A
multi-component S\'ersic or similar model asks how overlapping light
components combine to reproduce an image and provides quantities such as
component fluxes, effective radii, shapes, and S\'ersic indices
\citep{Peng2002,Peng2010,Erwin2015}. Centre-conditioned radial profiles ask a
different question: where does the observed field change relative to the
centres? They do not first assign the light to overlapping physical
components. An analytic law can subsequently be fitted to a
centre-conditioned profile when useful, but it is not required to define the
radial coordinate. The two approaches are therefore complementary rather than
competing descriptions of structure.

The same radial ordering can also support familiar structural summaries. For
a non-negative flux field, cumulative light along either coordinate can define
centre-conditioned light quantiles and concentration analogues, while slopes,
extrema, shoulders, breaks, and truncation locations can be stated relative
to an individual centre. These quantities would need to be calibrated for
particular applications and should not be assumed to replace measurements
defined using conventional circular or elliptical radii.

Once the support and centres are fixed, the geometry is also independent of
the tracer being measured. The same radial coordinates can therefore be used
for registered maps of colour, emission-line intensity or equivalent width,
stellar-population properties, dust attenuation, or line-of-sight kinematics.
This makes it possible to compare resolved changes at common
centre-conditioned radial positions across nuclei, clumps, stellar complexes,
or tidal structures.

The two JADES systems are intended as astronomical examples rather than as a
population study. They show that the distinction between relative
centre--boundary depth and normalized progression also appears in observed
non-axisymmetric light distributions. For these two systems, the centre
assignments and overall profile shapes remain stable under the support and
centre perturbations considered here. The same geometry can also be carried
between registered images, allowing different wavelengths or tracers to be
sampled at common radial positions without redefining the centres or source
support independently in each image. Physical interpretation of differences
between tracers would still require appropriate PSF matching, spatial
registration, and treatment of measurement uncertainties.

The construction remains conditional on the adopted connected support and
choice of centres. Changing the segmentation changes the admissible paths and
boundary distances, while changing the centres alters both the partition and
the radial ordering. The hard nearest-centre assignment also creates
interfaces between neighbouring centre-associated regions. Soft or
probabilistic assignments are possible extensions, but they would define a
different measurement. Finally, the coordinates describe projected
image-plane geometry. They do not establish physical association, determine
whether a centre is a nucleus or clump, or recover three-dimensional
distances.

Radial localization therefore need not be abandoned when a unique global
centre ceases to provide a useful description. Centre-conditioned radial
profiles recover the conventional normalized radius in the symmetric
single-centre limit while retaining centre identity and the radial location
of resolved features in irregular and multi-centred structures. They extend a
familiar astronomical measurement to systems in which a single radial axis
would mix structurally distinct regions, without requiring either a
parametric decomposition or a particular light-profile law.

\section*{Acknowledgements}

This work uses public data from the JWST Advanced Deep Extragalactic Survey
(JADES) and Galaxy Zoo: CANDELS. We thank the survey teams, catalogue
contributors, and Galaxy Zoo volunteers who made these data products publicly
available.

OpenAI Codex was used for assistance with code, figure preparation, and
language editing. The author determined the scientific scope, methodological
choices, validation strategy, interpretation, and conclusions of the work.

\section*{Data Availability}

The JWST/NIRCam imaging used in this work is publicly available through the
JADES DR2 data release \citep{Eisenstein2025}. The Galaxy Zoo: CANDELS
catalogue used to identify the astronomical examples is publicly available as
described by \citet{Simmons2017}.
The reference implementation of the radial construction is provided by
\textsf{RadialPaths}, available at
\href{https://rafaelsdesouza.com.br/radialpaths}
{\texttt{rafaelsdesouza.com.br/radialpaths}}.

\bibliographystyle{mnras}
\bibliography{references}
\appendix

\section{Interpretation and numerical evaluation of support-constrained distance}
\label{app:distance}

The support-constrained distance has a simple front-propagation
interpretation. Imagine launching a front from centre $c_k$ and allowing it to
move at unit speed while remaining inside the accepted support. In the
continuum, its arrival-time field $T_k(x)$ satisfies the Eikonal equation

\begin{equation}
 |\nabla T_k(x)| = 1,
 \qquad
 T_k(c_k)=0,
\end{equation}

with propagation restricted to $\Omega$. The arrival time at $x$ is then the
in-support geodesic distance,

\begin{equation}
 T_k(x)=d_G(c_k,x)=d_k(x).
\end{equation}

\begin{figure*}
 \centering
 \includegraphics[width=0.825\textwidth]{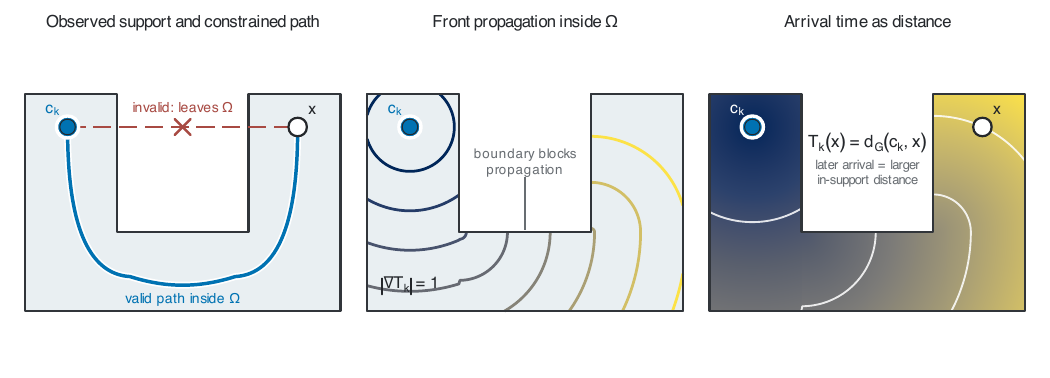}
 \caption{Front-propagation interpretation of support-constrained distance.
 Left: the straight-line path between a centre $c_k$ and a location $x$ may
 leave the observed support $\Omega$, whereas an admissible path remains
 inside it. Middle: equivalently, a front launched from $c_k$ propagates at
 unit speed within $\Omega$ and cannot cross its boundary. Right: the arrival
 time $T_k(x)$ defines the support-constrained distance,
 $T_k(x)=d_G(c_k,x)$. The illustration is schematic; numerical evaluation is
 described in the text.}
 \label{fig:frontprop}
\end{figure*}

This gives an intuitive meaning to the distance used throughout the paper. A
location is far from a centre when reaching it requires a long path through
the observed structure, even if the two points happen to lie close together
in the surrounding image plane. Folds, concave boundaries, internal excluded
regions, and narrow extensions can therefore increase the distance by
restricting the paths that remain inside the support. Figure~\ref{fig:frontprop}
shows this geometrically.

\begin{figure*}
 \centering
 \includegraphics[width=0.825\textwidth]{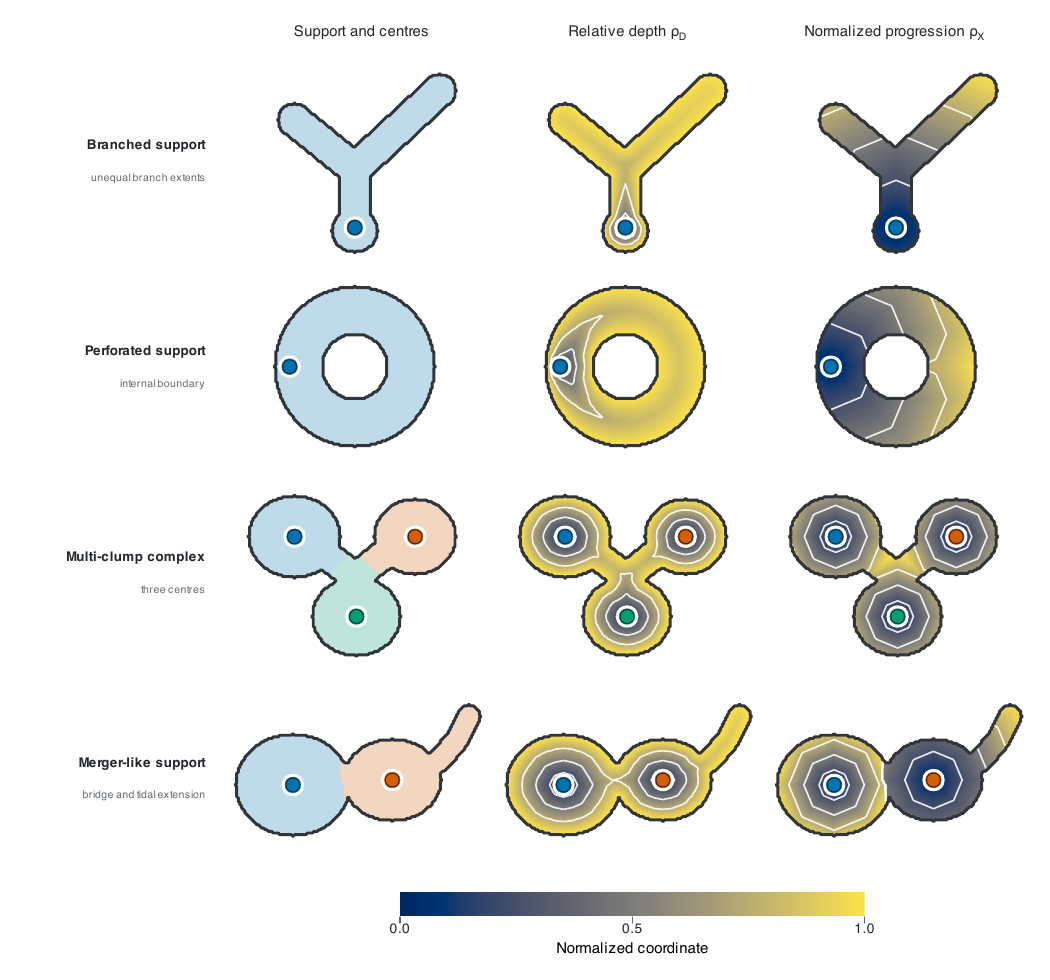}
 \caption{Additional geometries illustrating the complementary behaviour of
 relative centre--boundary depth $\rhoD$ and normalized progression $\rhoX$.
 Rows show, from top to bottom, a branched support with unequal branch
 extents, a perforated support containing an internal boundary, a
 three-centre clump complex, and an astronomy-inspired merger-like structure
 with a bridge and tidal extension. The left column shows the accepted
 support and centres; the middle and right columns show $\rhoD$ and $\rhoX$,
 respectively. In narrow branches, $\rhoD$ responds strongly to nearby
 lateral boundaries, whereas $\rhoX$ follows progression relative to the
 radial extent of the centre-associated region. In the perforated case, the
 internal hole is excluded from admissible paths and its edge contributes to
 the boundary distance. The multi-clump and merger-like examples show how the
 same construction extends to several centres and strongly non-convex
 astronomical morphologies.}
 \label{fig:additional_geometries}
\end{figure*}

The same continuum distance can be evaluated numerically in different ways.
For the synthetic information-retention example in
Fig.~\ref{fig:information}, we use a first-order fast-marching approximation
to the Eikonal equation \citep{Sethian1996}. On a pixel grid, we also use a
weighted neighbourhood graph in which accepted pixels are nodes and
shortest paths are restricted to remain inside the support
\citep{Dijkstra1959}. The JADES examples use an eight-neighbour graph with
unit axial steps and $\sqrt{2}$ diagonal steps. The additional geometries in
Fig.~\ref{fig:additional_geometries} use the same graph construction, so
excluded pixels cannot be traversed. This includes internal holes, whose
edges therefore contribute to the boundary-distance field.

These numerical schemes approximate the same continuum quantity $d_G$, but
they are not identical at finite pixel resolution. In particular, an
eight-neighbour graph retains the directional anisotropy of the underlying
pixel lattice. We quantify this effect using centred circular supports, for
which the continuum result is exactly

\begin{equation}
 \rhoD=\rhoX=\frac{r}{R}.
\end{equation}

For a circle with $R=128$ pixels, the 95th-percentile absolute deviations from
$r/R$ are 0.0165 for $\rhoD$ and 0.0524 for $\rhoX$ with the
eight-neighbour graph. The corresponding values for the first-order
fast-marching approximation are 0.0091 and 0.0042. Thus the equality with the
conventional normalized radius is exact for the continuum construction,
whereas a discrete image implementation retains a small dependence on the
pixel lattice.

The additional geometries in Fig.~\ref{fig:additional_geometries} show how
$\rhoD$ and $\rhoX$ respond to features that are absent from the circular
limit. The branched support introduces unequal extents, the perforated support
adds an internal boundary, the clump complex introduces several centres, and
the merger-like example combines a bridge with a tidal extension.

\begin{figure*}
 \centering
 \includegraphics[width=\textwidth]
   {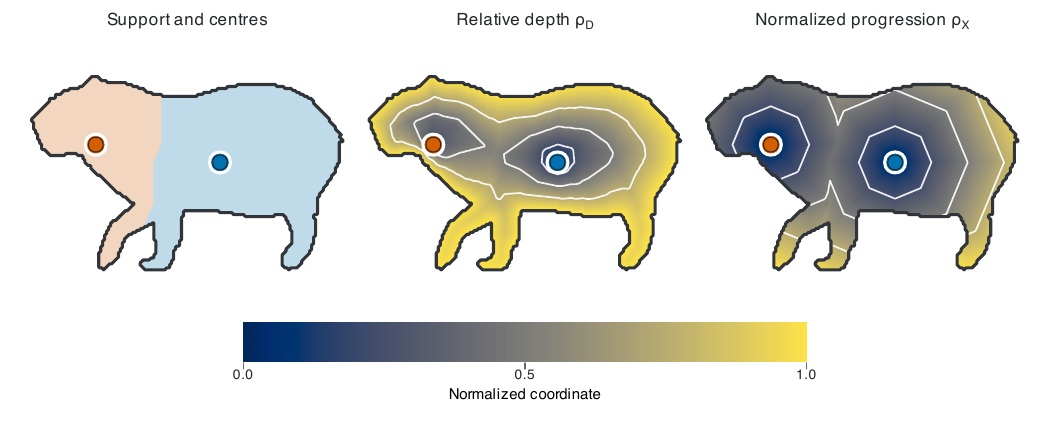}
 \caption{Radial coordinates on a capybara-shaped connected support. Left:
 centre-associated regions defined by two centres. Middle: relative
 centre--boundary depth $\rhoD$. Right: normalized progression $\rhoX$.
 White contours mark coordinate levels 0.25, 0.50, and 0.75. The support
 combines unequal extents, narrow connections, and an irregular boundary
 within a single outline, complementing the controlled geometries in
 Fig.~\ref{fig:additional_geometries}.}
 \label{fig:capybara}
\end{figure*}

The examples in Fig.~\ref{fig:additional_geometries} isolate these geometric
effects one at a time. Figure~\ref{fig:capybara} combines several of them in a
single irregular outline. The capybara-shaped support contains unequal
extents, narrow connections, and strongly varying boundary geometry, allowing
the two coordinates to be viewed together in a less idealized shape.

\label{lastpage}
\end{document}